\documentclass[runningheads]{llncs}
\usepackage{graphicx}
\usepackage{booktabs}
\usepackage[bookmarks=true%
,bookmarksnumbered=true%
,hypertexnames=false%
,breaklinks=true%
,colorlinks=true%
,linkcolor=blue%
,citecolor=blue%
,urlcolor=blue%
]{hyperref}
\hypersetup{
  pdfauthor = {},
  pdftitle = {},
  pdfsubject = {},
  pdfcreator = {LaTeX with hyperref package},
  pdfproducer = {pdflatex}
}
\usepackage{subfig}
\usepackage{multirow}
\usepackage{array}
\usepackage{color}
\usepackage{amsmath}
\usepackage[font={scriptsize}]{caption}

\newcolumntype{P}[1]{>{\centering\arraybackslash}p{#1}}

\begin{document}
%
\title{Identifying Notable News Stories}

\author{Antonia Saravanou\inst{1}\thanks{This work was done whilst interning at Bloomberg.} \and Giorgio Stefanoni\inst{2} \and Edgar Meij\inst{3}}

\institute{National and Kapodistrian University of Athens, Athens, Greece\\
\email{antoniasar@di.uoa.gr}
\and
Bloomberg, London, United Kingdom\\
\email{giorgio.stefanoni@gmail.com}
\and
Bloomberg, London, United Kingdom \\
\email{emeij@bloomberg.net}
}
\maketitle              
\begin{abstract}
The volume of news content has increased significantly in recent years and systems to process and deliver this information in an automated fashion at scale are becoming increasingly prevalent. 
One critical component that is required in such systems is a method to automatically determine how notable a certain news story is, in order to prioritize these stories during delivery. 
One way to do so is to compare each story in a stream of news stories to a notable event. 
In other words, the problem of detecting notable news can be defined as a ranking task; given a trusted source of notable events and a stream of candidate news stories, we aim to answer the question: ``Which of the candidate news stories is most similar to the notable one?''. 
We employ different combinations of features and learning to rank (LTR) models  and gather relevance labels using crowdsourcing. 
In our approach, we use structured representations of candidate news stories (triples) and we link them to corresponding entities. 
Our evaluation shows that the features in our proposed method outperform standard ranking methods, and that the trained model generalizes well to unseen news stories.

\end{abstract}

\section{Introduction}
\label{sec:intro}

With the rise in popularity of social media and the increase in citizen journalism, news is increasing in volume and coverage all around the world. 
As a result, news consumers run the risk of either being overwhelmed due to the sheer amount of news being produced, or missing out on news stories due to heavy filtering. 
To deal with the information overload, it is crucial to develop systems that can filter the noise in an intelligent fashion. 
Due to the highly condensed language used in news, automated systems have been developed to process them and generate well-defined structured representations from their content~\cite{tabari}. 
Each structure is a so-called triple that represents an event in the form of \textit{who} did \textit{what} to \textit{whom}, with additional metadata information about \textit{when} and \textit{where} this happened. 
Such representations (triples) form a \textit{knowledge graph} (KG). 
There are multiple computational benefits when searching, labeling, and processing KGs due to their clean and simple structure~\cite{yang2016fast,voskarides-etal-2015-learning}.

A common approach to measure notability of a news event is to track it through a proxy metric. For example, Naseri \textit{et al.}~\cite{Naseri:2019:APN:3331184.3331301} decide whether an article describes a notable event by counting the user interactions, while Setty \textit{et al.}~\cite{setty} cluster together similar news articles and then use the cluster size to decide if the common theme is notable. 
Wang \textit{et al.}~\cite{Wang:2018:DDK:3178876.3186175} propose a recommendation framework that takes as input a stream of news and predicts the user's click-through rate.

In this paper, we approach the problem of identifying notable news stories as a ranking task, i.e., we rank structured news stories represented 
as triples against notable events. We use \textit{Wikipedia's Current Events Portal} (WCEP)~\cite{wcep_bib}
as curated notable events and, using a combination of textual and semantic features, we
build a learning to rank (LTR) model to solve the ranking problem. 

\begin{table}[t]
	\caption{Example of a query $q_0$ and two candidate events $c_0$ and $c_1$.}
	\centering
	\scriptsize
	{	
		\label{tbl:wiki_example}	
		\scriptsize
		\begin{tabular}{p{5.3cm}p{0.2cm}p{6.5cm}}
			\toprule
			{\textbf{Query} $q_0$} & & {\textbf{Tagged Query}}\\
			\midrule
			{A suicide bomber detonates a vehicle full of explosives at a military camp in Gao, Mali, killing at least 76 people and wounding scores more in Mali's deadliest terrorist attack in history. \textit{Date}: 17 January 2017} & & {A [\textsc{Wiki}: \href{https://en.wikipedia.org/wiki/Suicide_attack}{Suicide\_attack}] detonates a [\textsc{Wiki}: \href{https://en.wikipedia.org/wiki/Vehicle}{Vehicle}] full of [\textsc{Wiki}: \href{https://en.wikipedia.org/wiki/Explosive}{Explosive}] at a military camp in [\textsc{Wiki}: \href{https://en.wikipedia.org/wiki/Gao}{Gao}], [\textsc{Wiki}: \href{https://en.wikipedia.org/wiki/Mali}{Mali}], [\textsc{Wiki}: \href{https://en.wikipedia.org/wiki/Murder}{Murder}] at least 76 people and wounding scores more in Mali's deadliest [\textsc{Wiki}: \href{https://en.wikipedia.org/wiki/Terrorism}{Terrorism}] in history. \textit{Date}: 17 January 2017}\\
			\midrule
		\end{tabular}
	}
	\centering
	{
		\label{tbl:icews_example}
		\label{tbl:example}

		\begin{tabular}{p{0.5cm}p{1.9cm}p{3.6cm}p{1.9cm}p{1.8cm}p{2cm}}
			{} & {\textbf{Subject}} & {\textbf{Predicate}} & {\textbf{Object}} & {\textbf{Date}} & {\textbf{Location}} \\
			\midrule
			{$c_0$} & {Armed Gang} & {Carry out suicide bombing} & {Armed rebel} & {17 Jan. 2017} & {Gao, Mali} \\			{$c_1$} & {Armed Gang} & {Carry out suicide bombing} & {Military} & {17 Jan. 2017} & {Bamako, Mali} \\
			
			\bottomrule
		\end{tabular}
	}
\end{table}

\section{Problem Statement}
\label{sec:problem_statement}

Let $\mathcal{Q} = [q_0, \dots, q_{k}]$ denote a stream of events, where each \textit{query event} $q_i\in\mathcal{Q}$ is a notable event composed of a textual description and of a publication date. 
Let $\mathcal{C} = [c_0, \dots, c_{l}]$ denote a stream of \textit{candidate} events. 
Each $c_j \in \mathcal{C}$ is a structured representation of a news story that consists of a \textit{triple} of the form $(s, p, o)$, where $s$ is the subject, $p$ is the predicate, and $o$ is the object, together with information about the \textit{location (city, country)} and the \textit{date} of the news story. 

Given a query $q_i \in \mathcal{Q}$ and a stream of candidates $\mathcal{C}$, we aim to rank each candidate $c_j \in \mathcal{C}$ by its relevance to the query $q_i$. 
A pair $(q_i, c_j)$ is considered as \textit{very relevant} when the information from $q_i$ and $c_j$ matches completely; it is considered as \textit{relevant} when some of the information matches; otherwise, it is considered as \textit{not relevant}. Table~\ref{tbl:icews_example} shows a query $q_0$ and two candidates $c_0$ and $c_1$. The pair $(q_0, c_0)$ is very relevant because $c_0$ matches $q_0$ completely; in contrast, the pair $(q_0, c_1)$ is relevant because $q_0$ and $c_1$ disagree only on the location of the event.

\section{Method}
\label{sec:method}
In this section we present our method to identify notable news stories which consists of three steps: (1) creating (query, candidate) pairs, (2) extracting textual and semantic features, and (3) training a learning to rank (LTR) model.

\noindent\paragraph{\textbf{(1) Creating Pairs.}}
\label{sec:create}
We create the set of all possible (query, candidate) pairs where (i) the query and the candidate have the same publication date, and (ii) the query and the candidate have at least {one word} in common as a pair is unlikely to be relevant if they share no words.

\noindent\paragraph{\textbf{(2) Extracting Features.}}
\label{sec:features}

We extract a set of features for each constructed pair. Our features can be classified into three groups as follows. 

\noindent{\textbf{(i) \textit{Features related to a component.}}} 
We compute the size of the query or the candidate (i.e., the number of terms in the query/candidate).

\noindent{\textbf{(ii) \textit{Features related to the pair.}}} 
We calculate the Okapi \textit{BM25} score, 
the term frequency (\textit{TF}) and the term frequency--inverse document frequency (\textit{TF--IDF}) for the query/candidate in the pair. 
We calculate these scores using the stemmed versions of the query/candidate (using the Porter Stemmer~\cite{porterstemmer}). 
We further define a similarity score, \textit{element match},
${EM}(q_i, ele_{c_j}) =  |q_i \cap ele_{c_j}| / |ele_{c_j}|$
where an element $ele_{c_j}$ is one of the: subject, predicate, object, description of the predicate, location, and the date in the candidate $c_j$.
For each of those, we calculate the fraction of the number of common terms between the element $ele_{c_j}$ and the query $q_i$ to the total number of terms in $ele_{c_j}$. 
%
%
In addition, we compute all \textit{EM} scores using the stemmed versions of the pair components. 
We also extract similarity scores for combinations of elements, as for example $EM$($q_i$, \textit{subject} $\cap$ \textit{predicate} $\cap$ \textit{object}) and $EM$($q_i$, \textit{city} $\cap$ \textit{country}).

\noindent{\textbf{(iii) \textit{Semantic features.}}} 
An entity is a well-defined, meaningful and unique way to characterize a word/phrase.
%
%
We therefore apply entity linking using the TagMe API~\cite{tagme} to identify entities (an example is shown in Table~\ref{tbl:wiki_example}). 
Given the tagged query and the tagged candidate, we calculate the number of common entities using the Jaccard similarity.

\paragraph{\textbf{(3) Ranking Pairs.}}
\label{sec:rank_pairs}
We then use our features to train a learning to rank model in order to obtain a ranking of pairs.
%
More details on the training and the tuning can be found in Section~\ref{sec:setup}.

\section{Experimental Setup}
\label{sec:setup}

\label{sec:dataset}

For the candidate news stories, we use the Integrated Crisis Early Warning System (ICEWS)~\cite{icews_bib} 
dataset which contains events that are automatically extracted from news articles using TABARI~\cite{icews,tabari}. 
This system uses 
grammatical parsing to identify events (\textit{who} did \textit{what} to \textit{whom}, \textit{when} and \textit{where}) using human-generated rules.
The events are triples consisting of coded actions between socio-political actors. The actors refer to individuals, groups, sectors and nation states. 
The actions are coded into $312$ categories. 
Geographical and temporal metadata are also associated with each triple (examples are shown in Table~\ref{tbl:icews_example}).

In our experiments, we use the same two weeks of data from ICEWS and WCEP. 
We remove triples with the generic action type ``Make statement'' as they do not convey any meaningful information. 
We then create pairs as described in Section~\ref{sec:create}. 
We build a crowdsourcing task (see below) to get golden truth labels. 
From the resulting annotated dataset, we only keep queries with at least one relevant ICEWS triple as there are, e.g., sports events in the WCEP dataset but not in the ICEWS dataset. 
In total, the resulting dataset contains $9.1$K pairs; $74$ queries and $123$ candidates per query on average. 
To evaluate our method in a real-world setting we split the dataset by date and use the first ten days for training, the next two days for validation, and the last two for testing.

\begin{table}[t]
	\centering
	\scriptsize
	\caption{Distribution of the relevance labels in the dataset.}
	\label{tbl:relevance_distribution}
	\begin{tabular}{p{2.7cm}p{1.8cm}p{1.8cm}p{2cm}p{1.8cm}}
		\toprule
		{} & \textbf{Train} & \textbf{Validate} & \textbf{Test} & \textbf{Total} \\
		
		\midrule
		{\textbf{Very Relevant}} & {220} ({4}\%) & {73} ({4}\%) & {47} ({3}\%) & {340} ({3}\%) \\
		{\textbf{Relevant}} & {106} ({2}\%)  & {20} ({1}\%) & {9} ({1}\%) & {135} ({1}\%) \\
		
		\textbf{Not Relevant} & {5219} ({94}\%) & {1959} ({95}\%) & {1475} ({96}\%) & {8653} ({96}\%) \\
		\bottomrule
	\end{tabular}
\end{table}

\noindent \textit{\textbf{Golden Truth.}} We employ crowdsourcing on the Figure-eight platform and ask annotators to judge the relevance of each pair on a 3-point scale (very relevant, relevant, not relevant).\footnote{\url{https://www.figure-eight.com/}}
Each pair $(q_i, c_j)$ is annotated by at least three annotators and we use majority voting to obtain the gold labels.
Our task obtains a inter-annotator agreement of $96.57$\%. 
Table~\ref{tbl:relevance_distribution} shows the distribution of relevance labels among pairs. 
The resulting dataset is highly skewed; with $3$\% annotated as \textit{very relevant}, $1$\% as \textit{relevant}, and $96$\% as \textit{not relevant}.

\noindent \textit{\textbf{Models.}} We explore various LTR algorithms and include results from RankBoost (RB)~\cite{rankboost}, lambdaMART (LM)~\cite{lambdaMART}, and Random Forest (RF)~\cite{rf}. 
We experiment using different sets of features: \textit{all} features (\textsc{All}), \textit{all} \textit{except entity-related} features (\textsc{All}$^{-}$), \textit{selected} features (\textsc{Sel}) and \textit{baseline} features (\textsc{B}). 
\textsc{Sel} features include BM25 and TF--IDF scores calculated from the original/stemmed versions, $EM$ scores for subject, predicate, object and location, and the number of entities in common and Jaccard similarity between the query and the candidate.
For \textsc{B} features, we only consider BM25 and TF--IDF scores calculated from the original/stemmed versions of the query and the candidate.
We evaluate using MAP, Precision@$k$, NDCG@$k$ and MRR.

\section{Results and Discussion}
\label{sec:discussion}

In this section we discuss our experimental results and answer the following research questions.
%
How does our method compare against the baselines? 
Does the performance vary with different parameter settings? 
Does the number of \textit{relevant} pairs affect performance? 
Do we benefit from tagging entities?

\subsection{Overall Performance}
\label{sec:overall_performance}

\begin{table}[t] \centering
	{
		\scriptsize
		\caption{Main results of the LTR models on our dataset.} \label{tbl:overall}
		\begin{tabular}{p{1.2cm}P{1.0cm}P{0.9cm}P{1.2cm}P{1.8cm}P{1.9cm}P{1.3cm}}
			\toprule
			{\textbf{}} & {\textbf{MAP}} & {\textbf{P@5}} & {\textbf{P@10}} & {{\textbf{NDCG}\textbf{@5}}} & {{\textbf{NDCG}\textbf{@10}}} & {\textbf{MRR}}\\
			\midrule
			{\textbf{RB$_{\textsc{All}}$}} & {0.53} & {0.42} & {0.3} & {0.6} & {\textbf{0.62}} & {\textbf{0.75}} \\
			{\textbf{LM$_{\textsc{All}}$}} & {0.44} & {0.38} & {0.31} & {0.51} & {0.56} & {0.65} \\
			{\textbf{RF$_{\textsc{All}}$}} & {\textbf{0.56}} & {\textbf{0.47}} & {{0.32}} & {\textbf{0.64}} & {0.61} & {\textbf{0.75}}\\
			\midrule
			
			{\textbf{RB$_{\textsc{B}}$}} & {0.37} & {0.33} & {0.29} & {0.37} & {0.45} & {0.6} \\
			{\textbf{LM$_{\textsc{B}}$}} & {0.34} & {0.31} & {0.29} & {0.36} & {0.44} & {0.6} \\
			{\textbf{RF$_{\textsc{B}}$}} & {{0.44}} & {{0.4}} & {\textbf{0.33}} & {{0.42}} & {{0.57}} & {{0.62}} \\
			\bottomrule
		\end{tabular}
	}
	
\end{table}

We compare the three LTR models on the \textsc{All} and \textsc{B} feature sets and show the results in Table~\ref{tbl:overall}. 
Our method (using \textsc{All} features) achieves better results than using just the baseline \textsc{B} features. 
For each model and feature set, we only show the best tuned model on the validation set. 
Our method consistently outperforms all baselines, achieving 5--12\% improvements on NDCG@10. 
These improvements are statistically significant with $p\leq 0.01$ using paired t-test. 
\begin{figure} [t]
	\centering%
	\caption{
    	\textbf{(Left)} 
			Results for each model on the validation set. 
			Each box shows the median and upper/lower quartiles. 
		\textbf{(Right)} 
			Performance using RB with selected features on two datasets.}
	\label{fig:exps}
	{\includegraphics[width=.67\linewidth]{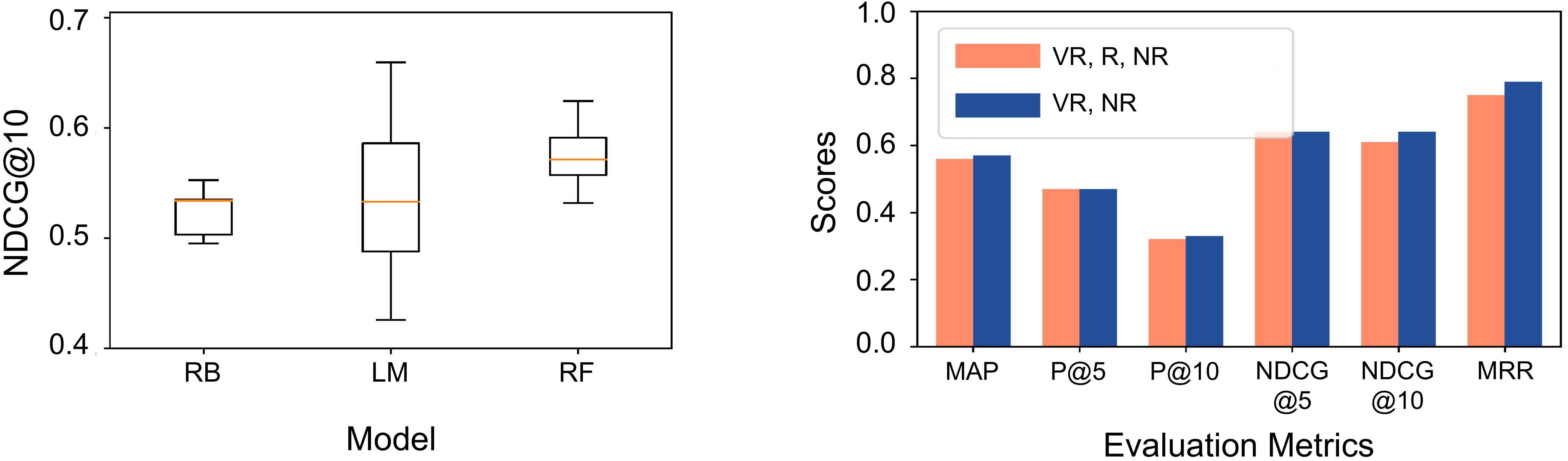}}

\end{figure}
%
%
%
%
%

%
%
%
%
%
We tune the parameters for each model on the validation set using NDCG@10. 
Figure~\ref{fig:exps} (left) shows the performance quartile plot using different parameter settings. 
RB and RF models show less sensitivity in the parameters tuning compared to LM. 
We evaluate the models when ranking pairs using all annotations (\textit{VR}, \textit{R}, and \textit{NR}). 
We perform the same experiment using only the \textit{VR} and \textit{NR} labeled pairs. 
Figure~\ref{fig:exps} (right) shows that the model achieves better results when excluding the \textit{R} labeled pairs. 
This is expected as the relevant label is very rare (only $1$\%, see Table~\ref{tbl:relevance_distribution}) and the models tend to consider it as noise.

Our next step is to evaluate different combinations of features (${\textsc{All}, \textsc{All}^-, \textsc{Sel}}$, \textsc{B}). 
We show our findings in Table~\ref{tbl:results}. 
First, we compare our method using \textsc{All$^-$} and \textsc{B} feature sets. 
We show that using the proposed features (Section~\ref{sec:features}) we achieve better performance for all LTR models. 
Second, we evaluate the performance of the models when we add the entity features by comparing \textsc{All} and \textsc{All$^-$}. 
In Table~\ref{tbl:results}, we show that there is a statistically significant improvement ($p\leq$ 0.01) on MRR (+7\%) when we add the entity--related features.

\begin{table}[t]
	\centering
	\caption{Results using binary relevance labels.}\label{tbl:results}
	{
		\scriptsize
		\centering
		\begin{tabular}{p{1.1cm}P{1.1cm}P{1.1cm}P{1.1cm}P{1.4cm}P{1.4cm}P{1.3cm}}
			\toprule
			{\textbf{}} & {\textbf{MAP}} & {\textbf{P@5}} & {\textbf{P@10}} & {{\textbf{NDCG@5}}} & {\textbf{NDCG@10}} & {\textbf{MRR}}\\
			\midrule
			{\textbf{RB$_{\textsc{All}}$}} & {\textbf{0.57}} & {0.42} & {0.3} & {0.61} & {\textbf{0.65}} & {0.69} \\
			{\textbf{LM$_{\textsc{All}}$}} & {0.53} & {0.4} & {0.3} & {0.56} & {0.61} & {0.71} \\
			{\textbf{RF$_{\textsc{All}}$}} & {\textbf{0.57}} & {\textbf{0.47}} & {\textbf{0.33}} & {\textbf{0.64}} & {0.64} & {\textbf{0.79}} \\ 
			\midrule
			
			{\textbf{RB$_{\textsc{All}^-}$}} & {0.52} & {\textbf{0.47}} & {\textbf{0.3}} & {0.62} & {0.62} & {0.68} \\
			{\textbf{LM$_{\textsc{All}^-}$}} & {0.44} & {0.33} & {0.28} & {0.47} & {0.54} & {0.65} \\
			{\textbf{RF$_{\textsc{All}^-}$}} & {\textbf{0.53}} & {0.44} & {\textbf{0.3}} & {\textbf{0.64}} & {\textbf{0.65}} & {\textbf{0.72}} \\ 
			\midrule
			
			{\textbf{RB$_{\textsc{Sel}}$}} & {\textbf{0.61}} & {0.44} & {0.28} & {\textbf{0.67}} & {\textbf{0.67}} & {\textbf{0.81}} \\
			{\textbf{LM$_{\textsc{Sel}}$}} & {0.53} & {0.4} & {0.27} & {0.56} & {0.6} & {0.75} \\
			{\textbf{RF$_{\textsc{Sel}}$}} & {0.55} & {\textbf{0.47}} & {\textbf{0.31}} & {0.62} & {0.65} & {0.62} \\
			\midrule
			
			{\textbf{RB$_{\textsc{B}}$}} & {\textbf{0.44}} & {\textbf{0.38}} & {0.28} & {\textbf{0.47}} & {0.51} & {0.6} \\
			{\textbf{LM$_{\textsc{B}}$}} & {0.38} & {0.33} & {0.28} & {0.39} & {0.47} & {0.54} \\
			{\textbf{RF$_{\textsc{B}}$}} & {0.42} & {0.31} & {\textbf{0.3}} & {0.42} & {\textbf{0.58}} & {\textbf{0.63}} \\
			\bottomrule
		\end{tabular}
	}
\end{table}

\subsection{Analysis}
\label{sec:analysis}
In this section, we show examples of the output from our best performing setting, i.e., RF with \textsc{All} features using the \textit{VR} and \textit{NR} labeled pairs. 
We show our best and worst per--query NDCG@10 performance. 
The best one achieves a score of 1, which indicates that our method was able to rank all pairs in the right order. 
The top--1 ranked pair is the query ``\textit{At least 15 children are killed and 45 more are injured after a school bus collides with a truck in Etah, India. Date: 20 Jan. 2017}'' and the candidate $<$\textit{Attacker (from India)}, \textit{Kill by physical assault}, \textit{Children (from India)}$>$ with metadata $<$\textit{Etah}, \textit{India}, \textit{20 Jan. 2017}$>$.
The item with the worst per--query NDCG@10 performance is ``\textit{Mexican drug lord Joaquin Guzman is extradited to the USA, where he will face charges for his role as leader of the Sinaloa Cartel. Date: 20 Jan. 2017}'' paired with the candidate $<$\textit{USA}, \textit{Host a visit}, \textit{Narendra Modi}$>$ with metadata $<$-, \textit{USA}, \textit{20 Jan. 2017}$>$. 
This query is about the extradition of a drug lord, while the candidate is about a visit of the Prime Minister of India. 
However, among the top--10 ranked candidates, the most relevant one is the triple $<$\textit{USA}, \textit{Arrest, detain, or charge with legal action}, \textit{Men (from Mexico)}$>$ with metadata $<$\textit{Kansas City}, \textit{USA}, \textit{20 Jan. 2017}$>$, ranked 9th. 
This shows that even in the worst ranking per--query, our method ranks a relevant candidate in the top--10.

In summary, we provide quantitative 
and qualitative 
performance analyses of our proposed method and we conclude that learning to rank is a viable method to determine notability of news stories. Among the key steps of our method are: (i) the extraction of textual and semantic features,  
and (ii) the exclusion of the pairs that do not convey strong signal, i.e., the ones labeled as `\textit{relevant}'. 
The \textsc{RF} model outperforms all baselines  
and it is also more robust with respect to hyperparameter settings. 
These findings show that our approach to detect notable news through ranking is a promising one. Although our method obtains high performance (MRR = 81\%), we believe we can attain further improvements by leveraging relations of the identified entities to discover implicitly relevant ones, such as $<$\textit{Narendra\_Modi}, \textit{isPrimeMinisterOf}, \textit{India}$>$.

\section{Conclusion and Future Work}
In this paper, we present a method to rank notable news representations which leverages textual and semantic features. 
Our evaluation on labeled pairs from WCEP and the ICEWS shows that our method is effective. 
In the future, we intend to include features based on the relations of the tagged entities from external KGs, such as DBPedia and Freebase.

%
%
%
\bibliographystyle{splncs04}
\bibliography{refs}
\end{document}